\documentclass[journal=jacsat,manuscript=article]{achemso}
\usepackage{chemformula} % Formula subscripts using \ch{}
\usepackage[T1]{fontenc} % Use modern font encodings

\usepackage{graphicx}% Include figure files
\usepackage{dcolumn}% Align table columns on decimal point
\usepackage{bm}% bold math
\usepackage{color}
\def\be{\begin{eqnarray}}
\def\ee{\end{eqnarray}}

\def\eps{\epsilon}
\def\veps{\varepsilon}
\def\epsE{\epsilon_{\rm E}}
\def\({\left(}
\def\){\right)}
\def\d{{\rm d}}
\def\e{{\rm e}}
\def\k10{k_{1\rightarrow 0}}
\newcommand{\olsi}[1]{\,\overline{\!{#1}}} % overline short italic

\author{Klavs Hansen}
\affiliation{Lanzhou Center for Theoretical Physics, 
Key Laboratory of Theoretical Physics of Gansu Province, 
Lanzhou University, Lanzhou, Gansu 730000, China}
\affiliation{Center for Joint Quantum Studies and Department of Physics, 
School of Science, Tianjin University, 92 Weijin Road, 
Tianjin 300072, China}
\author{Ori Licht}
\affiliation{Department of Physics and Insitute for Nanotechnology 
and Advanced materials, Bar Ilan University, Ramat-Gan 5290002, Israel}
\author{Adeliya Kurbanov}
\affiliation{Department of Physics and Insitute for Nanotechnology 
and Advanced materials, Bar Ilan University, Ramat-Gan 5290002, Israel}
\author{Yoni Toker}
\affiliation{Department of Physics and Insitute for Nanotechnology 
and Advanced materials, Bar Ilan University, Ramat-Gan 5290002, Israel}

\title{Cascade infrared thermal photon emission}

\begin{document}

\begin{abstract}
The time development of the excitation energy of molecules and 
clusters cooling by emission of thermal
vibrational infrared radiation has been studied.
The energy distributions and the photon
emission rates develop into  near-universal functions that can be characterized with only a few parameters,
irrespective of the precise vibrational spectra and oscillator strengths of the systems. 
The photon emission constant and emitted power averaged over all thermally 
populated states vary linearly with total excitation energy with a small offset. 
The time developments of ensemble
internal energy distributions are calculated with respect to their first two moments. 
For the derived linear dependence of the emission rate constant, these results are exact.
\end{abstract}

\section{Introduction}

Emission of photons is an important cooling channel for molecules and clusters in the absence of collisions. The decay channel is becoming  increasingly interesting in astrophysical context due to the growing 
number of molecules observed in interstellar space \cite{ISM_Mol_Herbst2009,ISM_Mol_Census2018}. 
The vastly different frequency factors and activation energies 
associated with the different thermal processes (fragmentation, thermionic emission, recurrent fluorescence, and infrared (IR) emission) create a hierarchy of time
scales on which the possible thermal decay channels of thermally excited molecules proceed.
The IR photon emission is located at the
lowest energies of this hierarchy, making
it the energy-loss channel that will
dominate at the latest stage of the 
cooling process.
The development of cryogenic storage rings has opened the possibility to study cooling processes on very long time scales in the laboratory
\cite{CTF_RC_AlClusters_2012,
Jap_RC_C7m_2014,CTF_RC_Co4m2018,DESIREE_Cn_RadiativeCooling,DESIREE_Cn_RadiativeCooling2,DESIREE_PAH_RC2020}. 
The extension of feasible storage times to thousands of seconds in these
devices places a special emphasis on the understanding of this long time emission of thermal infrared photons from stored  ions \cite{GotoJCP2013,Jap_RC_C7m_2014,DESIREE_Cn_RadiativeCooling,Ferrari2019,DESIREE_Cn_RadiativeCooling2,DESIREE_PAH_RC2020,IidaPRA2021,CSR_Al4_2022}

Before embarking on such a calculation of rate constants and energy distributions, it is worthwhile to consider two aspects of the thermal description of the vibrational degrees of freedom of a molecule or cluster.
One is the question of ergodicity.
The low internal energies of the emitting
molecules/clusters raise this question
with more force than for other decay
channels that need higher excitation energies to be sustained.
Harmonic oscillators do not couple, and
when a collection of rigorously uncoupled
harmonic oscillators is used to describe
the thermal properties of the vibrational
degrees of freedom, each of the modes will cool down 
independently; clearly, for a system of this type the cooling proceeds independently for all oscillators. 
The time dependence and average values can be calculated for such a system 
without any further ado. 

However, as molecular vibrations are not harmonic, energy can flow between different vibrational modes, a process know as intramolecular
vibrational redistribution (IVR) \cite{IVR_Bixon1968,IVR_Gruebele2004,
IVR_Makarov2012}.
IVR is indeed an essential component of our 
understanding of many phenomena at the heart
of physical chemistry, such as reaction rate
theories, fluorescence and coherent control.

The time scales associated with IVR are
determined by the degree of anharmonicity,
and scale with the density of states in the
simplest large molecule case (see ref.
\cite{NesbittJPC1996} for an introduction to
the subject). 
These IVR time scales are short enough to make IVR efficient on the timescales of milliseconds and longer, which are the  characteristic times of IR radiation. Radiative cooling measurements in storage rings do indeed indicate that IVR is the correct framework to describe the radiation at long times \cite{DESIREE_Cn_RadiativeCooling,DESIREE_Cn_RadiativeCooling2,CSR_Al4_2022}. The 'rapid exchange limit' denotes the case in which IVR is much faster than IR emission, and consequently, all states with the same total internal energy will mix rapidly. We will use the rapid exchange limit Ansatz here.

Another point worth mentioning is the little considered fact that vibrational motion carries angular momentum. 
The simplest case of linear molecules has been analyzed in \cite{HansenCPL2021}.
It is also well-known from the description of the excitation modes of helium droplets \cite{BrinkZPD1990,LehmannJCP2004,HansenPRB2007}.
The  conservation of this quantity imposes a constraint on the states that mix in the IVR process.
This clearly has consequences also for the photon emission rates.
However, a general theory is not yet available for molecules, and an in-depth  treatment of the subject is beyond the scope of this article. 
We will restrict the work here to  consider only the case of unrestricted IVR.

Within this framework we will show that vibrational cooling produces an almost linear scaling of the cooling rate constants with the internal energy of the system and, as a consequence, that the photon emission rate constant decays exponentially with time and with a rate constant that can be computed. 
We will use this scaling to calculate the time development 
of the general case where energy distributions are not delta functions.

\section{The harmonic cascade model}

We describe the $s=3N-6$ nuclear vibrations of a molecule with $N$ atoms as harmonic oscillators with energies  $\eps_i$, $i=1,...,s$. 
In this approximation, which is sufficiently accurate at the low energies we consider, the total excitation energy, $\varepsilon$, of a  molecule is therefore a sum over the mode energies:
\begin{equation}
    \varepsilon=\sum_i \nu_i \eps_i
\end{equation}
where $\nu_i$ are the number of excitation quanta in mode $i$.
A list of these occupation numbers, $\{ \nu_i \}$ specifies the system and will be denoted its state. 
Within the harmonic approximation, IR radiative emission only occurs between neighboring levels of each oscillator, i.e, $\nu_i\rightarrow \nu_i-1$. 
We will denote the rate constant for this transition, $k_{\nu_i\rightarrow \nu_i-1}^i=1/\tau_{\nu_i\rightarrow \nu_i-1}^i$, by the term emission rate constant of the state $\nu_i$ of oscillator $i$.
The matrix element that determines this rate constant for a particular vibrational mode, $i$, is the dipole matrix element:
\begin{equation}
\langle \nu_i-1|x|\nu_i \rangle_i \propto \sqrt{\nu_i}
\end{equation}
The rate constant for this particular mode is then given by:
\begin{equation}
    k_{\nu_i\rightarrow \nu_i-1}^i=\nu_i k^i_{1\rightarrow 0}.
\end{equation}
In the case of a single vibrational mode, with quantum energy $\eps_1$, the photon emission rate constant is therefore proportional to energy; $k(\veps)= \(\veps/\eps_1\) k_{1\rightarrow 0}^1$. 

\subsection{A Two Mode Example} 

\begin{figure*}[t]	
 \includegraphics[width=10cm]{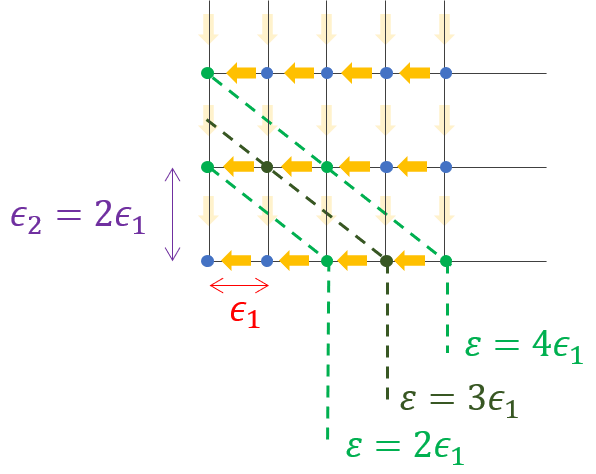}
   \caption{\label{TwoVibrationExample}An illustration of the harmonic cascade model for the case of a system with two modes for which $\eps_2=2\eps_1$, and only the low energy mode optically active. The allowed quantum states, defined by $\{ \nu_1, \nu_2 \}$, are 
    represented by points on a two dimensional lattice. 
    States with the same total energy $\veps=\nu_1\eps_1+\nu_2\eps_2$ lie on diagonals, as shown. IVR allows for the system to freely transit between states of the same energy. The yellow arrows 
    indicate the transitions that can occur via radiative cooling.}
\end{figure*}

Before treating the general case it is 
instructive to examine the simple case of a system with only two 
vibrational modes in which $\eps_2=2\eps_1$ and only one of the modes is IR active.
We set $k_{ \rightarrow 0}^2=0$. 
In this case, illustrated in 
Fig. \ref{TwoVibrationExample}, the 
number of states of the system where $\varepsilon=2m \eps_1$ or
$\varepsilon=2m \eps_1+1$ are both equal 
equal to $m+1$, where $m$ is a non-negative integer. 
For $\veps=2\eps_1$ the system is half the time in the state $\{ 2,0 \}$  which decays at a rate of $k_{2\rightarrow 1}^1=2k_{1\rightarrow 0}^1$, 
and half the time in the state $\{ 0,1 \}$ which decays at the rate of $k_{1 \rightarrow 0}^2=0$, and therefore: $k(2\eps_1)=k_{1\rightarrow 0}^1$. It is not hard to show that $k(2m\eps_1)$ and $k(2m \eps_1+1)$ are both 
equal to $mk_{1\rightarrow 0}^1$. Thus, once again we see a general linear dependence of the cooling rate on internal energy, albeit with fluctuations around the general trend.

\subsection{General Formulation}

In more realistic situations where the different vibrational energies are not exact multiples of each other, the rate constant for emission from the thermally populated states of oscillator $i$ is still calculated by summing over the products of populations and the energy-specified rate constants. 
The populations are calculated with the level densities (densities of states). 
Let $\rho(\veps,\nu_i=m)$ denote the density of states of total energy $\veps$ for which the vibrational mode $i$ is occupied $m$ times. 
The probability for the vibrational mode to be occupied exactly $m$ times is then $\rho(\veps, \nu_i=m)/\rho(\veps)$. 
Hence the rate of decay through mode $i$ is given by:
\begin{equation} \label{E:K 1}
    k^i(\veps)=\frac{1}{\rho(\veps)} \sum_{m=1}^{\veps/\eps_i} 
    \rho(\veps, \nu_i=m) k_{m \rightarrow m-1}^i 
    =\frac{1}{\rho(\veps)} \sum_{m=1}^{\veps/\eps_i}m 
    \rho(\veps, \nu_i=m) k_{1 \rightarrow 0}^i,
\end{equation}
which can be simplified using:
\begin{equation} \label{E:K 2}
    \sum_{m=1}^{\veps/\eps_i} m \rho(\veps,\nu_i=m) 
    = \sum_{m=1}^{\veps/\eps_i} \rho(\veps,\nu_i\geq m) 
    =\sum_{m=1}^{\veps/\eps_i} \rho(\veps-m\eps_i).
\end{equation}
Here $\rho(\nu_i\geq m,\veps)$ denotes the number of states of total energy $\veps$ where mode $i$ is occupied at least $m$ times. To prove the first equality we note that every state appears once in the sum on the left multiplied by $m$ while it is counted $m$ times in the sum on the right of this equality sign. 
For the equation on the right we note that once the mode $i$ is occupied at least $m$ times there remains an amount of $\veps-m\eps_i$ energy to distribute within the system. Using Eq. \ref{E:K 1} with Eq. \ref{E:K 2} leads to:
\begin{equation}
\label{E:General k^i(e)}
    k^i(\veps)=\frac{1}{\rho(\veps)}
    \sum_{m=1}^{[\veps/\eps_i]}\rho(\veps-m\eps_i) 
    k_{1 \rightarrow 0}^i =  \frac{1}{\rho(\veps)} \sum_{m=0}^{[\veps/\eps_i]-1}
    \rho((m+\Delta m)\eps_i)k_{1 \rightarrow 0}^i.
\end{equation}
The precise upper limit of the summation is indicated here with by taking the integer part of the energy in units of $\veps_i$, $[\veps/\veps_i]$.
The number $\Delta m$ is the fractional part of $\veps/\eps_i$, $\Delta m \equiv \veps/\veps_i - [\veps/\veps_i]$.
The total cooling rate constant of the molecule is the sum of such contributions from each vibration:
\be \label{E:Harm cascade model}
k(\veps)=\sum_i k^i(\veps)
\ee
The use of Eqs. \ref{E:General k^i(e)} and Eq. \ref{E:Harm cascade model} along with the density of states is known as the 'harmonic cascade model.  \cite{VJ2014_RF_C6, DESIREE_Cn_RadiativeCooling}

\begin{figure*}[t]	
 \includegraphics[width=10cm]{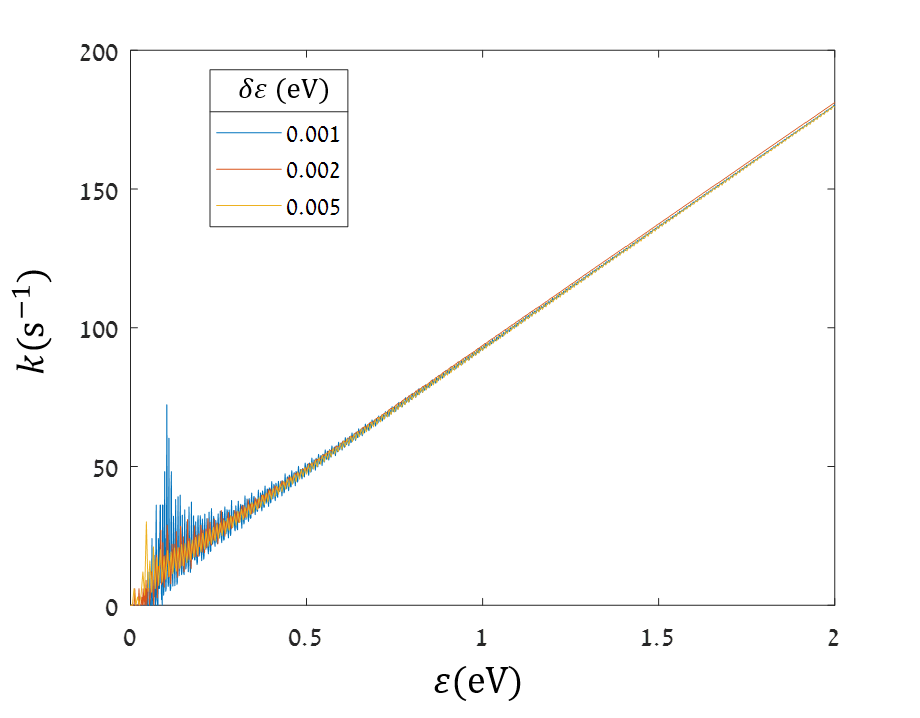}
    \caption{\label{F:K vs E dE dep} An example of the calculation of $k(\veps)$ for the case of Al$_4^-$, based on the 
    vibrational frequencies and Einstein coefficients calculated in Ref.\cite{CSR_Al4_2022}. }
\end{figure*}

An example of the results of such a calculation is shown in Fig. \ref{F:K vs E dE dep}. 
As with the single oscillator and the two mode examples discussed above, we observe that a general linear dependence of $k$ on $\veps$, as well as fluctuations on top of the general trend. 
The figure also shows the effect of a variation of the width of the energy bin, $\delta\veps$, used for the calculation of the density of states.
The curves for different values of $\delta \veps$ agree, up to the fluctuations on top of the general trend.
We need therefore not consider this issue any further. 

\section{Rate of Decay}

In the special case where all vibrational frequencies in a molecule are identical (the Einstein molecule), simple combinatorics can be used to show that $k(\veps)=\sum_i  k_{1\rightarrow 0}^i \veps/s\epsE=\langle k_{1\rightarrow 0} \rangle \veps/ \epsE$, with $\epsE$ the common vibrational quantum energy.
Trianglar brackets indicate averaging
over the different vibrational modes,
i.e.:
\begin{equation}
    \langle x \rangle \equiv \frac{1}{s}\sum_i x_i.
\end{equation}

To treat the general case in which the vibrational levels are not degenerate, we apply the criterion of agreement of the level density with the known high energy limit. 
Inserting the asymptotic limit for the density of states,
\be
\rho(\varepsilon) = \frac{\(\veps+ \frac{1}{2}\sum_{j=1}^s \epsilon_j \)^{s-1}}{(s-1)!\prod_{j=1}^s \eps_j}
\ee
into Eq. \ref{E:General k^i(e)} gives
\begin{eqnarray}\label{E:k vs e step 1}
    \frac{k^i(\veps)}{k^i_{1\rightarrow 0}}&=&\frac{\sum_{m=0}^{\veps/\eps_i-1}\left(m\eps_i + \frac{1}{2}\sum_i \eps_i\right)^{s-1}}{\left(\veps + \frac{1}{2}\sum_j \eps_j\right)^{s-1}} =\frac{\sum_{m=0}^{N_i-1}(m+V_i)^{s-1}}{(N_i+V_i)^{s-1}}\\\nonumber&=&\frac{1}{(N_i+V_i)^{s-1}}\left(\sum_{m=0}^{N_i+V-1}m^{s-1}-\sum_{m=0}^{V-1}m^{s-1}\right),
\end{eqnarray}
where $N_i\equiv \veps/\eps_i$ and $V_i \equiv \sum_j\eps_j/2\eps_i$. The $\sum_{m=0}^{V-1}m^{s-1}$ term is asymptotically negligible compared with the first and can therefore be ignored. The sum can be approximated using Faulhaber's formula (which are also the first two terms the Euler-Maclaurin formula):
\begin{equation}
    \sum_{m=0}^{N_i+V_i-1}m^{s-1}\simeq \frac{(N_i+V_i-1)^s}{s}+\frac{(N_i+V_i-1)^{s-1}}{2}\simeq \frac{(N_i+V_i)^s}{s}-\frac{(N_i+V_i)^{s-1}}{2}+O\left( (N_i+V_i)^{s-2} \right)
\end{equation}
Here the rightmost expression was derived using a Taylor expansion of the two powers. Inserting this approximation into Eq. \ref{E:k vs e step 1} leads to:
\begin{equation}\label{E:k vs e step 2}
    \frac{k^i(\veps)}{k^i_{1\rightarrow 0}} \simeq \frac{N_i+V_i}{s}-\frac{1}{2}=\frac{1}{s\eps_i}\veps-\frac{1}{2s\eps_i}\sum_j\eps_j-\frac{1}{2}
\end{equation}
which can be written as:
\begin{equation}
    k^i(\veps)=\frac{k^i_{1\rightarrow 0}}{s\eps_i}\veps-\frac{k^i_{1\rightarrow 0}}{2}\left(\frac{\langle \veps\rangle }{\eps_i}-1\right)\equiv a_i\veps+b_i
\end{equation}
\begin{equation}
    a_i=\frac{k_{1\rightarrow 0}^i}{\eps_i s}~~~~b_i=\frac{k^i_{1\rightarrow 0}}{2}\left(1-\frac{\langle \eps\rangle }{\eps_i}\right)
\end{equation}
An interesting consequence is that:
\begin{equation}
    \sum_i \eps_i k^i(\veps)=\frac{\sum k^i_{1\rightarrow 0}}{s}\veps=\langle k_{1\rightarrow 0}\rangle \veps
\end{equation}

\begin{figure*}[t]	
 \includegraphics[width=15cm]{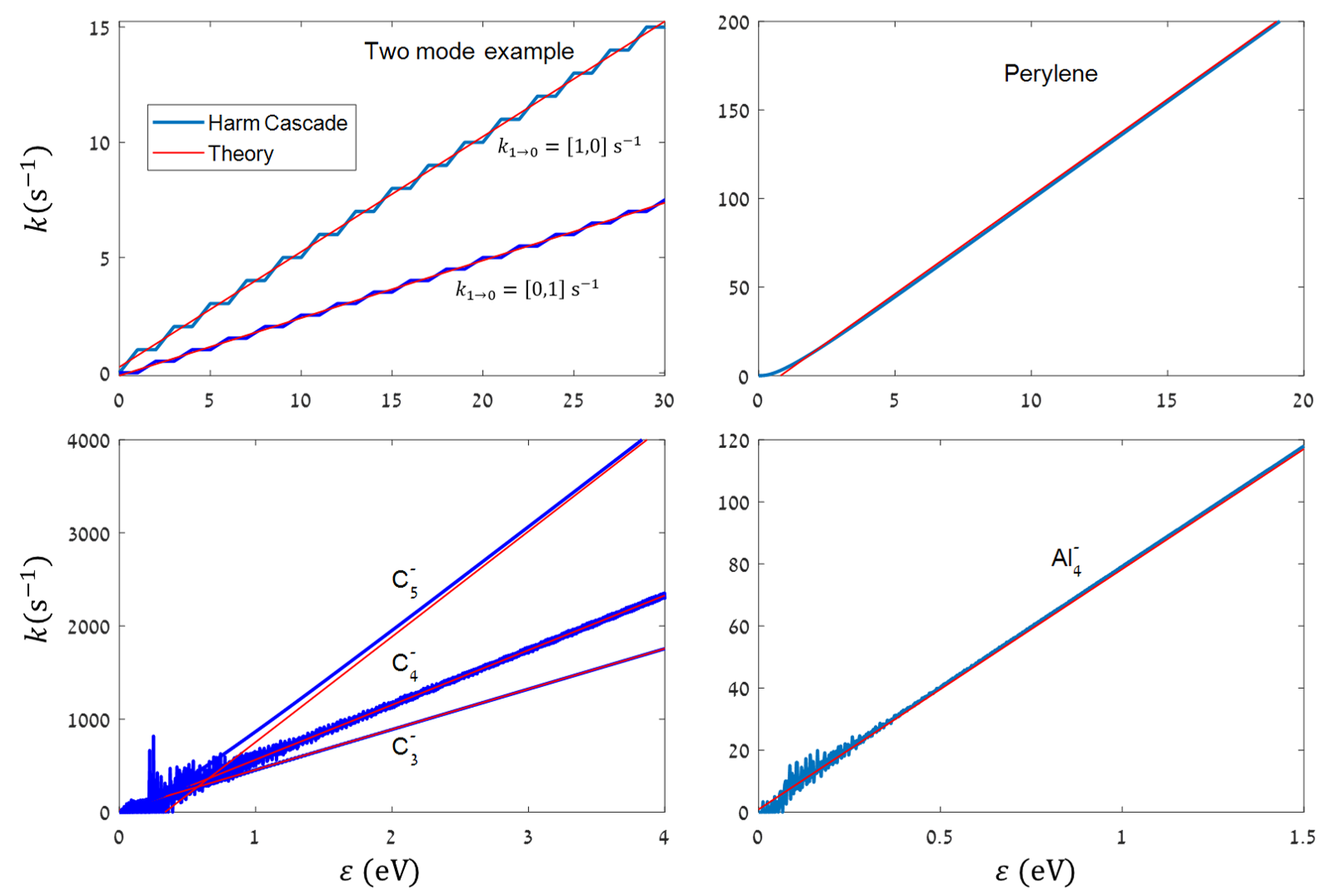}
    \caption{\label{F:K vs E} Calculation of the IR cooling rate $k(\varepsilon)$ using the harmonic cascade model (Blue), for a few representative cases.
    For the two-level model (Upper left panel) we calculated two cases, where either the first or the second vibration is IR active and the other is not. 
    For the cases of Perylene \cite{DESIREE_PAH_RC2020} (upper right frame), carbon clusters \cite{DESIREE_Cn_RadiativeCooling}(lower left frame) and Al$_4^-$ (lower right frame), we use the vibrational frequencies and Einstein coefficients calculated in the respective works. 
    The red line in each panel is the theoretical $k(\veps)$ calculated according to Eq. \ref{E:Harm cascade model}.}
\end{figure*}

The total decay rate $k(\veps)$ is therefore given by:
\begin{equation}\label{E:k vs eps theory}
k(\veps)=\sum_i k^i(\veps)\equiv a\veps+b
\end{equation}
with:
\begin{equation}
    a=\sum_i a_i=\left\langle \frac{k_{1\rightarrow 0}}{\eps}\right\rangle, ~~~ b=\sum_i b_i=\frac{s}{2}\left(\langle \k10 \rangle - \langle \frac{\k10}{\eps} \rangle \langle \eps \rangle \right)
\end{equation}
%\begin{equation}
%    b=\sum_i b_i=\sum_ik^i_{1\rightarrow 0}
%\end{equation}
In particular we find that, in the asymptotic limit, $k$ has a linear dependence on internal energy $\varepsilon$. 
A similar scaling has previously been given in Ref. \cite{TerzievaIJMS2000}, albeit without the offset in the linear dependence of the rate constant that will be derived here.
Figure \ref{F:K vs E} shows a comparison of the calculation of the decay rate $k(\varepsilon)$ using the harmonic cascade model calculated with the Beyer-Swinehart level density, compared with the results of Eq. \ref{E:k vs eps theory}.
Notably, while Eq. \ref{E:k vs eps theory} does not account for the fluctuations on top of the general trend, it does seem to capture the general trend extremely well.

\section{Population Evolution}

\begin{figure*}[t]	
 \includegraphics[width=15cm]{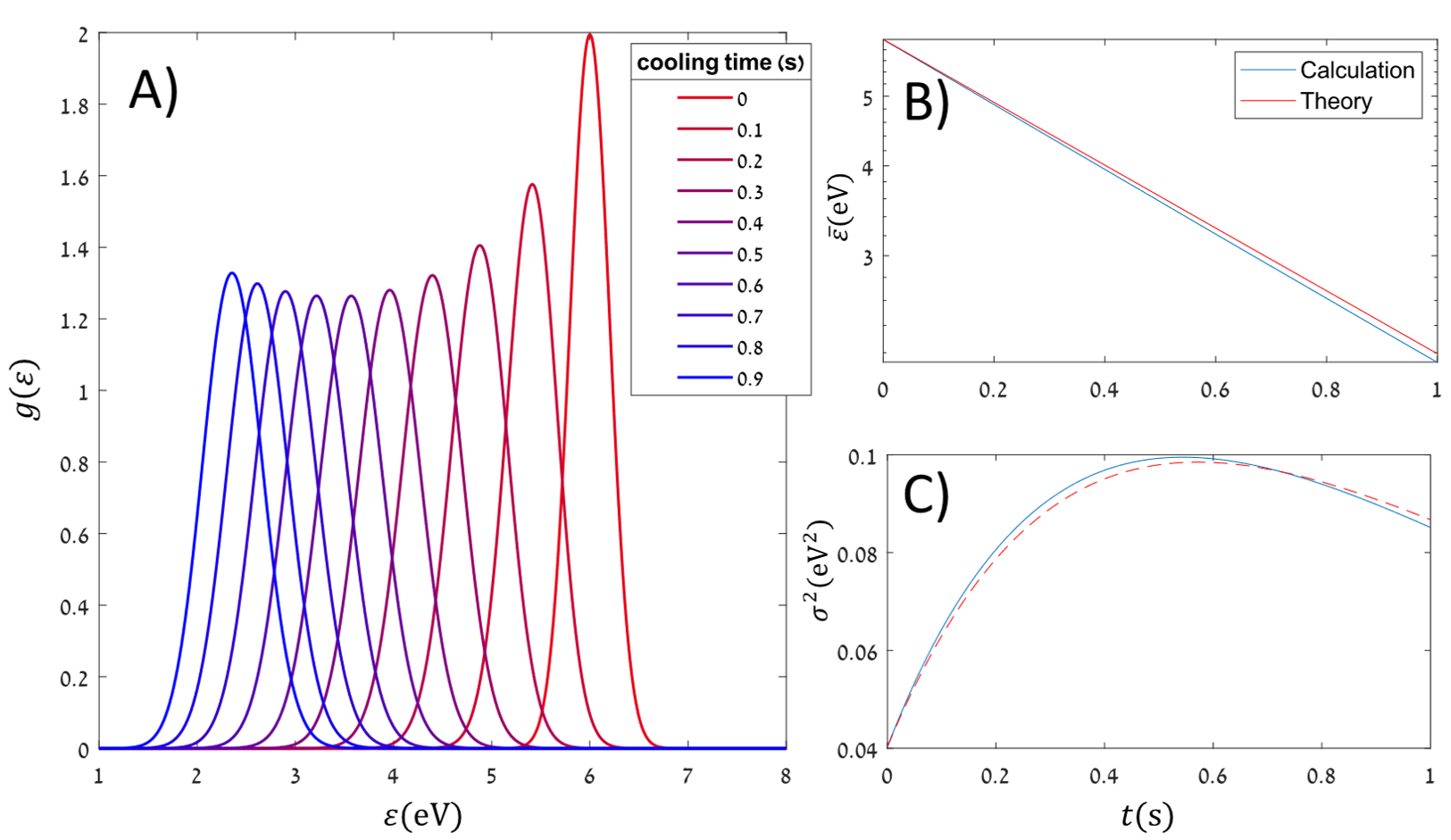}
    \caption{\label{F:Poplation evolution} A: calculation of the evolution $g(\varepsilon)$ according to the harmonic cascade model, for the case of Al$_4^-$, starting with an initial Gaussian distribution with a width of $0.2~$eV. B: mean energy, $\olsi{\veps}$ as a function of time, the red line corresponds to Eq. \ref{Eq:Mean Energy vs time}, C: the variance as a function of time, compared with Eq. \ref{Eq:Var vs time}.}
\end{figure*}

Let $g(\veps)$ denote the normalized energy distribution of a system,
\be
\int_0^\infty g(\veps)d\veps=1.
\ee
The harmonic cascade model allows a calculation of the evolution of $g(\veps)$ with time according to:
\begin{equation}\label{E:g evolution}
\frac{{\rm d} g(\veps)}{{\rm d} t}=\sum_i k^i(\veps+\eps_i)g(\veps+\eps_i)-k(\veps)g(\veps),
\end{equation}
where the sum runs over the vibrational modes. 
The first term comes from states with higher energy radiatively decaying into the states of interest, and $k(\veps) \equiv \sum_i k^i(\veps)$ is the total rate constant for decay out of the system with energy $\veps$. In this section we will find the time evolution of $g(\veps)$ in terms of its mean, $\olsi{\veps}=\int_0^\infty g(\veps)\veps d\veps$, and its variance: $\sigma^2=\int_0^\infty g(\veps)(\veps-\olsi{\veps})^2d\veps$. 
We will analyze the high energy limit, with the understanding that this limit is a fairly good approximation to reasonably low energies, up to the fluctuation induced by the quantum discreteness (see fig. \ref{F:K vs E}).
In this limit the total emission rate constant is given by Eq. \ref{E:k vs eps theory}. 
With that expression, the energy develops as:
\be
\frac{\d \veps}{\d t} = -\sum_i k^i (\veps)\eps_i = -\sum_i \eps_i  k^i_{1\rightarrow 0}
\(\frac{\veps}{s\eps_i} +
\frac{\sum_j\eps_j}{2s\eps_i}-
\frac{1}{2}\) = -\frac{\veps}{s} \sum_i k^i_{1\rightarrow 0}.
\ee
Note that upon summation the offset terms cancel out ($\sum\eps_i\left(\frac{\sum_j\eps_j}{2s\eps_i}-
\frac{1}{2}\right)=0$). 
The energy therefore decreases exponentially:
\be
\label{mean-t}
\veps(t) = \veps(0){\e}^{-\sum_i k^i_{1\rightarrow 0}t/s}=\veps(0){\e}^{-\langle \k10 \rangle t}.
\ee
The exponent in this expression is the same for all energies. 
Averaging over a distribution will therefore give the same decrease for the mean energy:
\be \label{Eq:Mean Energy vs time}
\olsi{\veps}=\olsi{\veps}(t=0){\e}^{-\langle \k10 \rangle t}.
\ee
%Defining a time dependent total rate %constant according to: $k(t)\equiv %k(\veps(t))$ leads to:
%\be
%k(t)= \sum_ik^i_{1\rightarrow 0}
%\(\frac{1}{s\eps_i}
%\veps(0)\e^{-\sum_i %k^i_{1\rightarrow 0}t/s}
%+ \frac{\sum_j\eps_j}{2s\eps_i}-%\frac{1}{2}\)
%\ee

The time evolution of the variance of $g(\veps)$, $\sigma^2$, is governed by the equation
\be
\frac{\d \sigma^2}{\d t} = 
\frac{\d}{\d t} \int_0^{\infty} (\veps - \olsi{\veps})^2g(\veps)\d \veps=
\int_0^{\infty} \left[
\frac{\d g}{\d t}\times \(\veps - \olsi{\veps}\)^2 -
2g(\veps)\times (\veps - \olsi{\veps})\frac{\d \olsi{\veps}}{\d t} \right] \d \veps.
\ee
The last term is zero by definition of the mean value. The first term requires the time evolution of $g$, according to Eq.\ref{E:g evolution}.
This gives:
\be
\frac{\d \sigma^2}{\d t} = 
\int_0^{\infty} \left[\(\veps - \olsi{\veps}\)^2 \(\sum_i k^i(\veps+\eps_i)g(\veps+\eps_i)-k^i(\veps)g(\veps)  \) \right] \d \veps,
\ee
or, exchanging integration and summation,
\be
\frac{\d \sigma^2}{\d t} = 
\sum_i \int_0^{\infty} \(\veps - \olsi{\veps}\)^2 \left[ k^i(\veps+\eps_i)g(\veps+\eps_i)-k^i(\veps)g(\veps)  \right] \d \veps.
\ee
The rate constants are represented with the two (known) parameters $a_i, b_i$ as
\be
k_i = a_i \veps + b_i.
\ee
Each term in the sum can then be written as
\be
\frac{\d \sigma_i^2}{\d t}&\equiv&\int_0^{\infty} \(\veps - \olsi{\veps}\)^2 \left[ \(a_i(\veps+\eps_i)+b_i\)g(\veps+\eps_i)-\(a_i(\veps)+b_i\)g(\veps)  \right] \d \veps = \\\nonumber
&-&\int_0^{\infty} 
\(\veps - \olsi{\veps}\)^2  \(a_i\veps+b_i\)g(\veps)  \d \veps\\\nonumber
&+&\int_0^{\infty} \(\veps- \olsi{\veps} + \eps_i\)^2 \(a_i(\veps+\eps_i)+b_i\)g(\veps+\eps_i)
\d \veps\\\nonumber
&+&\int_0^{\infty} 
\eps_i^2  \(a_i\veps+b_i\)g(\veps)  \d \veps\\\nonumber
&-&2\int_0^{\infty} \eps_i \(\veps+\eps_i - \olsi{\veps}\) \(a_i(\veps+\eps_i)+b_i\)g(\veps+\eps_i)
\d \veps.
\ee
The first and the second terms on the right hand side cancel.
Collecting terms from the remaining two integrals gives
\be
\frac{\d \sigma_i^2}{\d t} &=&
\int_0^{\infty} 
\eps_i^2 \(a_i\veps+b_i\)g(\veps)  \d \veps
-2\int_0^{\infty} \eps_i \(\veps + \eps_i- \olsi{\veps}\) \(a_i(\veps+\eps_i)+b_i\)g(\veps+\eps_i)
\d \veps\\\nonumber 
&=& a_i \eps_i^2 \olsi{\veps} +b_i\eps_i^2
-2a_i\eps_i\sigma^2.
\ee
Inserting the values of $a_i$ and $b_i$ gives
\be
\frac{\d \sigma_i^2}{\d t}
= a_i\eps_i\(\eps_i\olsi{\veps}
-2\sigma^2 \)+b_i\eps_i^2 
= \frac{k^i_{1\rightarrow 0}}{s}
\(\eps_i\olsi{\veps}-2\sigma^2 \)
+\frac{1}{2}
k^i_{1 \rightarrow 0}\veps_i^2-\frac{k^i_{1 \rightarrow 0}}{2} 
\eps_i \langle \eps \rangle.
\ee
Summation over all modes gives 
\be
\frac{\d \sigma^2}{\d t}
= \frac{\olsi{\veps}}{s} \sum_i k^i_{1 \rightarrow 0} \eps_i
-\sigma^2 \frac{2}{s}\sum_i 
k^i_{1 \rightarrow 0} + \frac{1}{2}\sum_i k^i_{1 \rightarrow 0}\eps_i^2 -
\sum_i \frac{k^i_{1 \rightarrow 0}}{2}\eps_i
\langle \eps \rangle,
\ee
or
\be
\frac{\d \sigma^2}{\d t}
= \olsi{\veps} \langle k\veps\rangle -2\langle k \rangle \sigma^2 +\frac{s}{2}\langle k \veps^2 \rangle-
\frac{s}{2}\langle k \veps\rangle 
\langle \eps \rangle.
\ee
This has the solution  
\be \label{Eq:Var vs time}
\label{sigma}
\sigma^2 &=&\frac{s}{4\langle k \rangle}
\(\langle k\veps^2 \rangle-\langle k\veps \rangle\langle \veps \rangle\) + \olsi{\veps}(0)\frac{\langle k\veps \rangle}{\langle k \rangle }
{\e}^{-\langle k\rangle t}\\\nonumber
&+&
\(\sigma^2(0) -\frac{s}{4\langle k \rangle}\(\langle k\veps^2 \rangle-\langle k\veps\rangle \langle \veps\rangle \)-\olsi{\eps}(0) \frac{\langle k\eps\rangle}{\langle k \rangle }\){\e}^{-2\langle k \rangle t}
\ee
For the special case where all $\eps_i = \eps_0$ are identical, the equation reduces to
\be
\sigma^2 = \olsi{\eps}(0)\eps_0 \e^{-\langle k \rangle t}\(1-\e^{-\langle k \rangle t}\)
+\sigma^2(0)\e^{-2 \langle k \rangle t}.
\ee
At long times the ratio of the standard deviation to the mean value for this case approaches the value
\be
\frac{\sigma}{\olsi{\eps}}
\approx 
\sqrt{\frac{\eps_0}{\olsi{\eps}(0)}}
\e^{\langle k \rangle t/2}.
\ee
Hence the relative widths of these distributions increase as the particle cools.

Another point worth considering is if the width can increase in absolute terms. 
This will happen if the initial distribution is sufficiently narrow. 
Taking the derivative at zero time shows that the condition for this is  
\be
\sigma(0) \leq \sqrt{\frac{\olsi{\eps}(0)\eps_0}{2}}.
\ee
For an ensemble with an initially canonical distribution and $s$ thermally activated  vibrational degrees of freedom, the value of $\sigma$ is $\sqrt{s}\eps$. 
This translates into the condition \be
\olsi{\eps}(0) \leq \frac{\eps_0}{2s}.
\ee
This is a very small width and is rarely of interest in realistic situations.

It is of interest to compare the results for the mean and variances with those that describe an emission with a continuum spectrum. 
These cases were treated in \cite{StenfalkEPJD2007} with a photon absorption cross section proportional to the photon energy to the power $n$.
Both the mean energy and the width varies asymptotically as a power law with the power $-1/(n+3)$, i.e. in parallel, in contrast to the exponential and non-parallel variations found here.

\section{Discussion and Summary}

The dynamics of vibrational cooling in the rapid exchange model has been analyzed, showing that the cooling rate has a linear dependence on internal energy. This implies that if one observes experimentally a deviation from the linear behavior, this means that the assumption of rapid IVR would be incorrect. \cite{CSR_Al4_2022} Furthermore, the consequences regarding the  temporal evolution of the means and the variances of energy distributions that are cooled by vibrational transitions have been studied. 

The main result is an exponential decrease of the mean energy with time. The time dependence of the width of the energy distribution involves several terms but will asymptotically approach an exponentially decreasing value, albeit with a time constant of twice of the one for the mean value. Distributions will therefore tend to broaden relative to the mean with time. 

The high energy results fail at excitation energies around one quantum per mode and less.
This is not surprising and is also clear from inspection of the literature results on C$_5^-$ \cite{GotoJCP2013}.
The energy distributions for this cluster decrease with a power of time, with the same power for width and mean. 
Simulated data reproduce these features with measured/calculated low energy and ground state properties, and the differences to the present results are therefore not due to new physics, but rather a consequence of the low energy limit. In lieu of a closed form expression for the photon emission rate constants rate in this limit, the equations \ref{E:General k^i(e)},
and \ref{E:g evolution} can be used to calculate the temporal development of energy distributions. 
\bibliography{Refs}

\providecommand{\latin}[1]{#1}
\makeatletter
\providecommand{\doi}
  {\begingroup\let\do\@makeother\dospecials
  \catcode`\{=1 \catcode`\}=2 \doi@aux}
\providecommand{\doi@aux}[1]{\endgroup\texttt{#1}}
\makeatother
\providecommand*\mcitethebibliography{\thebibliography}
\csname @ifundefined\endcsname{endmcitethebibliography}
  {\let\endmcitethebibliography\endthebibliography}{}
\begin{mcitethebibliography}{24}
\providecommand*\natexlab[1]{#1}
\providecommand*\mciteSetBstSublistMode[1]{}
\providecommand*\mciteSetBstMaxWidthForm[2]{}
\providecommand*\mciteBstWouldAddEndPuncttrue
  {\def\EndOfBibitem{\unskip.}}
\providecommand*\mciteBstWouldAddEndPunctfalse
  {\let\EndOfBibitem\relax}
\providecommand*\mciteSetBstMidEndSepPunct[3]{}
\providecommand*\mciteSetBstSublistLabelBeginEnd[3]{}
\providecommand*\EndOfBibitem{}
\mciteSetBstSublistMode{f}
\mciteSetBstMaxWidthForm{subitem}{(\alph{mcitesubitemcount})}
\mciteSetBstSublistLabelBeginEnd
  {\mcitemaxwidthsubitemform\space}
  {\relax}
  {\relax}

\bibitem[Herbst and Dishoeck(2009)Herbst, and Dishoeck]{ISM_Mol_Herbst2009}
Herbst,~E.; Dishoeck,~E. F.~V. Complex organic interstellar molecules.
  \emph{Ann. Rev. Astron. Astr.} \textbf{2009}, \emph{47}, 427--480\relax
\mciteBstWouldAddEndPuncttrue
\mciteSetBstMidEndSepPunct{\mcitedefaultmidpunct}
{\mcitedefaultendpunct}{\mcitedefaultseppunct}\relax
\EndOfBibitem
\bibitem[McGuire(2018)]{ISM_Mol_Census2018}
McGuire,~B.~A. 2018 Census of Interstellar, Circumstellar, Extragalactic,
  Protoplanetary Disk, and Exoplanetary Molecules. \emph{Astrophys. J. Suppl.
  S.} \textbf{2018}, \emph{239}, 17\relax
\mciteBstWouldAddEndPuncttrue
\mciteSetBstMidEndSepPunct{\mcitedefaultmidpunct}
{\mcitedefaultendpunct}{\mcitedefaultseppunct}\relax
\EndOfBibitem
\bibitem[Lange \latin{et~al.}(2012)Lange, Froese, Menk, Bing, Fellenberger,
  Grieser, Laux, Orlov, Repnow, Sieber, Toker, von Hahn, Wolf, and
  Blaum]{CTF_RC_AlClusters_2012}
Lange,~M.; Froese,~M.~W.; Menk,~S.; Bing,~D.; Fellenberger,~F.; Grieser,~M.;
  Laux,~F.; Orlov,~D.~A.; Repnow,~R.; Sieber,~T.; Toker,~Y.; von Hahn,~R.;
  Wolf,~A.; Blaum,~K. Radiative cooling of $Al_4^-$ and $Al_5^-$ in a cryogenic
  environment. \emph{New J. Phys.} \textbf{2012}, \emph{14}, 065007\relax
\mciteBstWouldAddEndPuncttrue
\mciteSetBstMidEndSepPunct{\mcitedefaultmidpunct}
{\mcitedefaultendpunct}{\mcitedefaultseppunct}\relax
\EndOfBibitem
\bibitem[Najafian \latin{et~al.}(2014)Najafian, Pettersson, Dynefors,
  Shiromaru, Matsumoto, Tanuma, Furukawa, Azuma, and Hansen]{Jap_RC_C7m_2014}
Najafian,~K.; Pettersson,~M.~S.; Dynefors,~B.; Shiromaru,~H.; Matsumoto,~J.;
  Tanuma,~H.; Furukawa,~T.; Azuma,~T.; Hansen,~K. Radiative cooling of $C_7^-$.
  \emph{J. Chem. Phys.} \textbf{2014}, \emph{140}, 104311\relax
\mciteBstWouldAddEndPuncttrue
\mciteSetBstMidEndSepPunct{\mcitedefaultmidpunct}
{\mcitedefaultendpunct}{\mcitedefaultseppunct}\relax
\EndOfBibitem
\bibitem[Breitenfeldt \latin{et~al.}(2018)Breitenfeldt, Blaum, George,
  G\"{o}ck, Guzm\'{a}n-Ram\'{\i}rez, Karthein, Kolling, Lange, Menk, Meyer,
  Mohrbach, Niedner-Schatteburg, Schwalm, Schweikhard, and
  Wolf]{CTF_RC_Co4m2018}
Breitenfeldt,~C.; Blaum,~K.; George,~S.; G\"{o}ck,~J.;
  Guzm\'{a}n-Ram\'{\i}rez,~G.; Karthein,~J.; Kolling,~T.; Lange,~M.; Menk,~S.;
  Meyer,~C.; Mohrbach,~J.; Niedner-Schatteburg,~G.; Schwalm,~D.;
  Schweikhard,~L.; Wolf,~A. Long-Term Monitoring of the Internal Energy
  Distribution of Isolated Cluster Systems. \emph{Phys. Rev. Lett.}
  \textbf{2018}, \emph{120}, 253001\relax
\mciteBstWouldAddEndPuncttrue
\mciteSetBstMidEndSepPunct{\mcitedefaultmidpunct}
{\mcitedefaultendpunct}{\mcitedefaultseppunct}\relax
\EndOfBibitem
\bibitem[Bull \latin{et~al.}(2019)Bull, Scholz, Carrascosa, Kristiansson,
  Eklund, Punnakayathil, de~Ruette, Zettergren, Schmidt, Cederquist, and
  Stockett]{DESIREE_Cn_RadiativeCooling}
Bull,~J.~N.; Scholz,~M.~S.; Carrascosa,~E.; Kristiansson,~M.~K.; Eklund,~G.;
  Punnakayathil,~N.; de~Ruette,~N.; Zettergren,~H.; Schmidt,~H.~T.;
  Cederquist,~H.; Stockett,~M.~H. Ultraslow radiative cooling of Cn- (n=3-5).
  \emph{J. Chem. Phys.} \textbf{2019}, \emph{151}, 114304\relax
\mciteBstWouldAddEndPuncttrue
\mciteSetBstMidEndSepPunct{\mcitedefaultmidpunct}
{\mcitedefaultendpunct}{\mcitedefaultseppunct}\relax
\EndOfBibitem
\bibitem[Stockett \latin{et~al.}(2020)Stockett, Bull, Buntine, Carrascosa,
  Anderson, Gatchell, Kaminska, Nascimento, Cederquist, T., and
  Zettergren]{DESIREE_Cn_RadiativeCooling2}
Stockett,~M.; Bull,~J.~N.; Buntine,~J.~T.; Carrascosa,~E.; Anderson,~E.~K.;
  Gatchell,~M.; Kaminska,~M.; Nascimento,~R.~F.; Cederquist,~H.; T.,~S.~H.;
  Zettergren,~H. Radiative cooling of carbon cluster anions C$_{2n+1}^-$ ($n=
  3-5$). \emph{Eur. Phys. J. D} \textbf{2020}, \emph{74}, 150\relax
\mciteBstWouldAddEndPuncttrue
\mciteSetBstMidEndSepPunct{\mcitedefaultmidpunct}
{\mcitedefaultendpunct}{\mcitedefaultseppunct}\relax
\EndOfBibitem
\bibitem[Stockett \latin{et~al.}(2020)Stockett, Bull, Buntine, Carrascosa, Ji,
  Kono, Schmidt, and Zettergren]{DESIREE_PAH_RC2020}
Stockett,~M.~H.; Bull,~J.~N.; Buntine,~J.~T.; Carrascosa,~E.; Ji,~M.; Kono,~N.;
  Schmidt,~H.~T.; Zettergren,~H. Unimolecular fragmentation and radiative
  cooling of isolated PAH ions: A quantitative study. \emph{J. Chem. Phys.}
  \textbf{2020}, \emph{153}, 154303\relax
\mciteBstWouldAddEndPuncttrue
\mciteSetBstMidEndSepPunct{\mcitedefaultmidpunct}
{\mcitedefaultendpunct}{\mcitedefaultseppunct}\relax
\EndOfBibitem
\bibitem[Goto \latin{et~al.}(2013)Goto, Sund\'{e}n, Shiromaru, Matsumoto,
  Tanuma, Azuma, and Hansen]{GotoJCP2013}
Goto,~M.; Sund\'{e}n,~A.; Shiromaru,~H.; Matsumoto,~J.; Tanuma,~H.; Azuma,~T.;
  Hansen,~K. {Direct observation of internal energy distributions of
  C$_5^-$}{}. \emph{J. Chem. Phys.} \textbf{2013}, \emph{139}, 054306\relax
\mciteBstWouldAddEndPuncttrue
\mciteSetBstMidEndSepPunct{\mcitedefaultmidpunct}
{\mcitedefaultendpunct}{\mcitedefaultseppunct}\relax
\EndOfBibitem
\bibitem[Ferrari \latin{et~al.}(2019)Ferrari, Janssens, Lievens, and
  Hansen]{Ferrari2019}
Ferrari,~P.; Janssens,~E.; Lievens,~P.; Hansen,~K. Radiative cooling of
  size-selected gas phase clusters. \emph{Int. Rev. Phys. Chem.} \textbf{2019},
  \emph{38}, 405\relax
\mciteBstWouldAddEndPuncttrue
\mciteSetBstMidEndSepPunct{\mcitedefaultmidpunct}
{\mcitedefaultendpunct}{\mcitedefaultseppunct}\relax
\EndOfBibitem
\bibitem[Iida \latin{et~al.}(2021)Iida, Kuma, Kuriyama, Furukawa, Kusunoki,
  Tanuma, Hansen, Shiromaru, and Azuma]{IidaPRA2021}
Iida,~S.; Kuma,~S.; Kuriyama,~M.; Furukawa,~T.; Kusunoki,~M.; Tanuma,~H.;
  Hansen,~K.; Shiromaru,~H.; Azuma,~T. {IR-photon quenching of delayed electron
  detachment from hot pentacene anions}. \emph{Phys. Rev. A} \textbf{2021},
  \emph{104}, 043114\relax
\mciteBstWouldAddEndPuncttrue
\mciteSetBstMidEndSepPunct{\mcitedefaultmidpunct}
{\mcitedefaultendpunct}{\mcitedefaultseppunct}\relax
\EndOfBibitem
\bibitem[D.~Muell()]{CSR_Al4_2022}
D.~Muell,~Y. T. e.~a.,~H.~Kreckel \textit{In preparation:} Radiative cooling of
  Al$_4^-$ and Al$_5^-$ in a cryogenic environment. \relax
\mciteBstWouldAddEndPunctfalse
\mciteSetBstMidEndSepPunct{\mcitedefaultmidpunct}
{}{\mcitedefaultseppunct}\relax
\EndOfBibitem
\bibitem[Bixon and Jortner(1968)Bixon, and Jortner]{IVR_Bixon1968}
Bixon,~M.; Jortner,~J. Intramolecular Radiationless Transitions. \emph{J. Chem.
  Phys.} \textbf{1968}, \emph{48}, 715\relax
\mciteBstWouldAddEndPuncttrue
\mciteSetBstMidEndSepPunct{\mcitedefaultmidpunct}
{\mcitedefaultendpunct}{\mcitedefaultseppunct}\relax
\EndOfBibitem
\bibitem[Gruebele and Wolynes(2004)Gruebele, and Wolynes]{IVR_Gruebele2004}
Gruebele,~M.; Wolynes,~P.~G. {Vibrational Energy Flow and Chemical Reactions}.
  \emph{Acc. Chem. Res.} \textbf{2004}, \emph{37}, 261--267\relax
\mciteBstWouldAddEndPuncttrue
\mciteSetBstMidEndSepPunct{\mcitedefaultmidpunct}
{\mcitedefaultendpunct}{\mcitedefaultseppunct}\relax
\EndOfBibitem
\bibitem[Makarov \latin{et~al.}(2012)Makarov, Malinovsky, and
  Ryabov]{IVR_Makarov2012}
Makarov,~A.~A.; Malinovsky,~A.~L.; Ryabov,~E.~A. Intramolecular vibrational
  redistribution: from high-resolution spectra to real-time dynamics.
  \emph{Uspekhi Fizicheskih Nauk} \textbf{2012}, \emph{182}, 977\relax
\mciteBstWouldAddEndPuncttrue
\mciteSetBstMidEndSepPunct{\mcitedefaultmidpunct}
{\mcitedefaultendpunct}{\mcitedefaultseppunct}\relax
\EndOfBibitem
\bibitem[Nesbitt and Field(1996)Nesbitt, and Field]{NesbittJPC1996}
Nesbitt,~D.~J.; Field,~R.~W. {Vibrational Energy Flow in Highly Excited
  Molecules: Role of Intramolecular Vibrational Redistribution}{}. \emph{J.
  Phys. Chem.} \textbf{1996}, \emph{100}, 12735--12756\relax
\mciteBstWouldAddEndPuncttrue
\mciteSetBstMidEndSepPunct{\mcitedefaultmidpunct}
{\mcitedefaultendpunct}{\mcitedefaultseppunct}\relax
\EndOfBibitem
\bibitem[Hansen and Ferrari(2021)Hansen, and Ferrari]{HansenCPL2021}
Hansen,~K.; Ferrari,~P. Vibrational angular momentum level densities of linear
  molecules. \emph{Chem. Phys. Lett.} \textbf{2021}, \emph{768}, 138385\relax
\mciteBstWouldAddEndPuncttrue
\mciteSetBstMidEndSepPunct{\mcitedefaultmidpunct}
{\mcitedefaultendpunct}{\mcitedefaultseppunct}\relax
\EndOfBibitem
\bibitem[Brink and Stringari(1990)Brink, and Stringari]{BrinkZPD1990}
Brink,~D.~M.; Stringari,~S. Density of states and evaporation rate of helium
  clusters. \emph{Z Phys D} \textbf{1990}, \emph{15}, 257--263\relax
\mciteBstWouldAddEndPuncttrue
\mciteSetBstMidEndSepPunct{\mcitedefaultmidpunct}
{\mcitedefaultendpunct}{\mcitedefaultseppunct}\relax
\EndOfBibitem
\bibitem[Lehmann(2004)]{LehmannJCP2004}
Lehmann,~K.~K. Bias in the temperature of helium nanodroplets measured by an
  embedded rotor. \emph{J. Chem. Phys.} \textbf{2004}, \emph{120}, 513\relax
\mciteBstWouldAddEndPuncttrue
\mciteSetBstMidEndSepPunct{\mcitedefaultmidpunct}
{\mcitedefaultendpunct}{\mcitedefaultseppunct}\relax
\EndOfBibitem
\bibitem[Hansen \latin{et~al.}(2007)Hansen, Johnson, and Kresin]{HansenPRB2007}
Hansen,~K.; Johnson,~M.~D.; Kresin,~V.~V. Density of states of helium droplets.
  \emph{Phys. Rev. B} \textbf{2007}, \emph{76}, 235424\relax
\mciteBstWouldAddEndPuncttrue
\mciteSetBstMidEndSepPunct{\mcitedefaultmidpunct}
{\mcitedefaultendpunct}{\mcitedefaultseppunct}\relax
\EndOfBibitem
\bibitem[Chandrasekaran \latin{et~al.}(2014)Chandrasekaran, Kafle, Prabhakaran,
  Heber, Rappaport, Rubinstein, Schwalm, Toker, and Zajfman]{VJ2014_RF_C6}
Chandrasekaran,~V.; Kafle,~B.; Prabhakaran,~A.; Heber,~O.; Rappaport,~M.;
  Rubinstein,~H.; Schwalm,~D.; Toker,~Y.; Zajfman,~D. Determination of Absolute
  Recurrent Fluorescence Rate Coefficients for C6–. \emph{J. Phys. Chem.
  Lett.} \textbf{2014}, \emph{5}, 4078--4082\relax
\mciteBstWouldAddEndPuncttrue
\mciteSetBstMidEndSepPunct{\mcitedefaultmidpunct}
{\mcitedefaultendpunct}{\mcitedefaultseppunct}\relax
\EndOfBibitem
\bibitem[Terzieva and Herbst(2000)Terzieva, and Herbst]{TerzievaIJMS2000}
Terzieva,~R.; Herbst,~E. Radiative electron attachment to small linear carbon
  clusters and its significance for the chemistry of diffuse interstellar
  clouds. \emph{Int. J. Mass Spetrom.} \textbf{2000}, \emph{201},
  135--142\relax
\mciteBstWouldAddEndPuncttrue
\mciteSetBstMidEndSepPunct{\mcitedefaultmidpunct}
{\mcitedefaultendpunct}{\mcitedefaultseppunct}\relax
\EndOfBibitem
\bibitem[Stenfalk and Hansen(2007)Stenfalk, and Hansen]{StenfalkEPJD2007}
Stenfalk,~J.; Hansen,~K. Energy distributions of clusters cooled by thermal
  radiation. \emph{Eur. Phys. J. D} \textbf{2007}, \emph{43}, 101\relax
\mciteBstWouldAddEndPuncttrue
\mciteSetBstMidEndSepPunct{\mcitedefaultmidpunct}
{\mcitedefaultendpunct}{\mcitedefaultseppunct}\relax
\EndOfBibitem
\end{mcitethebibliography}
\end{document}